\documentclass[11pt]{article}
\usepackage{graphicx}
\def\title{\begin{center}\Large\bf}
\def\author(s){\vspace{0.3cm}\large\rm}
\def\text{\end{center}}

\begin{document}






\title
Detection of high-frequency variability in chromospherically active
stars

\bigskip



\author(s)
B.E. Zhilyaev $^1$, M.V. Andreev $^2$, A.V. Sergeev $^2$

\bigskip

\smallskip

\noindent $^1$ {\small {\it Main Astronomical Observatory, NAS
of Ukraine,  27 Zabolotnoho, 03680 Kiev, Ukraine}} \\

\noindent {\small {\it e-mail:}} {\small{\bf zhilyaev@mao.kiev.ua}}

\smallskip

\noindent $^2${\small {\it International Centre for Astronomical,
Medical and Ecological Research \\ Terskol settlement,
Kabardino-Balkaria, 361605 Russia}} \\

\smallskip



\text


\section*{Abstract}

We have carried out high-speed photometry of three chromospherically
active stars, BD +15 3364, II Peg, and SAO 52355 with the Zeiss 2-m
telescope, as well as low-resolution spectroscopy of SAO 52355 with
the Zeiss-600 telescopes at Peak Terskol. BD +15 3364 is known as
chromospherically active star of the  spectral type G0. II Peg is
the RS CVn binary. It has a spectral type K2IV-Ve. SAO 52355 is a
field star of the spectral type K0 III. The X-ray observations taken
by Ginga and ROSAT indicate in II Peg and SAO 52355 coronae plasma
temperatures as high as $10^{7}$ K. These two stars are supposed to
have high-powered chromospheres. Photometric observations of all
three stars show high-frequency variations in brightness in the UBV
bands at subsecond range. Intensity variations are found peaked at
frequency around about 0.5 Hz, spanning the range up to 1.5 Hz for
BD +15 3364, II Peg, and up to about 35 Hz for SAO 52355. The
relative power of fluctuations reaches ($10^{-3.7} - 10^{-4.2}$) in
the UBV bands. Spectroscopic monitoring of SAO 52355 showed
variations of emission in the Balmer lines and in the CaII H, K
lines at time intervals ranging from seconds to minutes. From the
power spectrum data one can find that variations in the intensities
of the CaII H, K and $H_{\gamma}$ lines are 3.2\% and 1.5\%,
respectively. This allows us to assert the existence of intense
microflaring activity in these stars.
\vspace{0.5 cm}

\noindent {{\bf keywords}$\,\,\,\,$\large stars: late-type -- stars:
chromospheres -- methods: observational -- techniques: photometric}

\large

\section{Introduction}

In this paper we present a set of high time-resolution observations
of some chromospherically active stars and their reference stars in
the UBV bands acquired in August 2011 at the Peak Terskol. Here we
discuss photometric observations of three stars: BD +15 3364, II
Peg, and SAO 52355. BD +15 3364 is known as chromospherically young,
kinematically old star [5], spectral type G0, magnitudes in Johnson
V = 8.66, B = 9.27. II Peg is the RS CVn binary. According to the
GCVS its photometric magnitude spanning the range V:(7.18  - 7.78),
period of variation: 6.7026 days, Spectral type: K2IV-Ve. The X-ray
observations taken by Ginga indicate coronae plasma temperatures as
high as $10^{8}$ K [6]. SAO 52355 is a field star, spectral type: K0
III, Johnson magnitude V: 9.86, Johnson B-V color, computed from the
Tycho catalog B-V: 1.33.

We found on these three stars photometric evidence of chromospheric
activity, as this will be shown below. It is found that
chromospherically active stars show strong Hydrogen and Ca II
emission lines as well as high X-ray luminosity.  As mentioned by
[1], the high X-ray flux from chromospherically active stars
detected by space observatories could be explained by assuming that
the heating of its corona results from a large number of small
flares.  It is well-known that the solar corona is heated by the two
most favored agents involving magnetic fields, namely MHD waves and
transients, such as flares, micro- and nanoflares (see [3] and
references therein).  It is expected that when two
oppositely-directed magnetic fields come closer, the current density
of the contained plasma increases considerably, so that even a small
resistivity is quite sufficient to convert magnetic energy of plasma
to thermal energy via magnetic reconnection. The reconnect heating
by foot-point motions was proposed first by Parker [4]. Reconnection
is associated with direct magnetic field dissipations. A forest of
closed magnetic loops has foot points ankered in photospheric
regions. The mechanical energy flux is generated by foot point
motions. These motions increase the energy stored in the entwined
magnetic field. This system can return to a minimum energy
configuration only after a reconnection (or a cascade of
reconnections). It is thought that these small and frequent
reconnection events give rise to the microflare heating [2].

Our primary goal is to present clear evidence of intense
microflaring activity in stars of such a type, using a high-speed
photometer that can operate in three bands (UBV) with sampling
frequencies up to 100 Hz. In the subsequent sections we describe the
observation, then the analysis technique applied and the main
results.


\section{Observations}

\noindent  The observational data for program stars and their
references stars were obtained over 3 nights in August 2011. We used
the 2-m Ritchey-Chretien telescope at Peak Terskol (North Caucasus,
3100 m a.s.l.) with a high-speed two-channel UBVR photometer [9].
The integration time was 0.01 and 0.1 s. Hence, our instrumentation
allows to detect signals with frequencies up to 50 Hz.

A new promising approach based on the theory of count statistics
allowed us to detect activity in some chromospherically active stars
and late-type giants on short time-scales. The actual value of
variability caused by atmospheric scintillation was determined from
measurements of a reference star. The intrinsic activity of the star
was found as a difference between the observed relative power and
that of the atmospheric scintillations.

The SAO 52355 spectra were observed with the grating spectrograph on
the Zeiss-600 telescope at Peak Terskol on May 30-31, 2010. A blazed
transmission grating is included in the converging beam in the
Zeiss-600 filter wheel. The observations were acquired with a
SpectraVideoTM Camera, Pixel Vision, Inc. 2. The grating
spectrograph observations cover the wavelength range from $\approx
3700\,{\AA}$ to $9000\,{\AA}$ and a time interval of 1600 seconds.
The exposure time was 8 seconds. The wavelength scale after
calibration is accurate to about $30\,{\AA}$. The grating spectrum
has a resolution of R $ \approx 100$ at $4800\,{\AA}$.


\section{Detection and estimation of high-frequency variability}

\noindent A high-frequency variability is present, for example, in
all dwarf novae in all stages of activity and other cataclysmic
variables [7] in a form of random brightness fluctuations, with a
continuous distribution in the corresponding frequency domain. The
frequency distribution gives some indication about the geometrical
extent of the variable source, namely there is either a point-like
source or an extended, optically thick one. In view of random nature
of variability, there is some difficulty in detection of any
intrinsic low-amplitude fluctuations close to the noise level. A new
promising tool for solving the problem relies on the theory of count
statistics. The intrinsic activity can be detected using the
factorial moments [8]
\begin{equation}\label{Klauder1}
    \langle \frac{n!}{(n-k)!}\rangle = \langle (n(n-1)...(n-k+1)
    \rangle = n_{[k]}
\end{equation}
where $n$ is the count rate, the angle brackets denote time
averaging. It is convenient to use the normalized factorial moments
\begin{equation}\label{Klauder2}
    h_{[k]} = \frac{n_{[k]}}{\langle n  \rangle ^{k}}
\end{equation}
In the case of a Poisson statistics it can be shown that, for any
$k$, $h_{[k]} \equiv 1$. Hence, any significant deviation of
$h_{[k]}$ from one may signal the presence of variability.
Attempting to detect activity on shortest time-scales, we applied
this relatively straightforward approach. The expression for the
factorial moment of the second order
\begin{equation}\label{Klauder3}
    \varepsilon = \frac{\sigma^{2}-<n>}{<n>^{2}} = h_{[2]}-1
\end{equation}
specifies the relative power of fluctuations $\varepsilon$ in the
frequency range $\Delta f = (\frac{1}{2\Delta t} - \frac{1}{m\Delta
t})$, where $\Delta t$ is the sampling time, $m$ the length of the
data segment, $<n>$ and $\sigma^{2}$ are the sample mean count rate
and the variance, respectively. Choosing appropriate values of $m$
and $\Delta t$, one can calculate the power spectrum of fluctuations
by averaging $\varepsilon $ over time. In our case, we adopt $m$ =
5. The standard deviation for $h_{[2]}$, as shown in [10] is defined
by the relation
\begin{equation}\label{Klauder4}
    std(h_{[2]})\simeq \frac{1}{<n>}\sqrt{\frac{1}{N}}
\end{equation}
The actual value of $\varepsilon$ caused by atmospheric
scintillation can be determined from measurements of a reference
star. The difference in $\varepsilon$ between the observed relative
power of the star and that of the atmospheric scintillations is
taken to be the intrinsic $\varepsilon$ - spectrum of star.

\section{Results and discussion}
\subsection{SAO 52355}

\noindent In photometry of bright stars scintillation noise is
usually a dominant error source. The intensity of scintillations is
log-normally distributed. Hence,for weak scintillations in bright
stars the standard errors are identical, if expressed in the
stellar-magnitude or relative power scales. It is important to use a
nearby comparison star to evaluate the scintillation component of
the covariance in Eqn. (3) and exclude the atmospheric effect.

Fig. 1 shows for SAO 52355 and the comparison star the powers of
variation and standard errors in the B band. The estimated
scintillation powers do not depend on stellar magnitudes and agree
within the limits of experimental errors up to 5 Hz. It is
interesting to note that situation outside of the (5 - 35) Hz range
significantly differs due to intrinsic activity of SAO 52355 in the
B band.

Fig. 2 shows the difference in $\varepsilon$ between the observed
relative power of the variable star and that of the atmospheric
scintillations taken from the $\varepsilon$ - spectrum of a
reference star. Thus, Fig. 2 indicates the presence of intrinsic
activity in the range $(\sim 5 - 35)$ Hz in SAO 52355. The power
excess reaches $\sim 7.8\cdot10^{-5}$ of the total power in the B
band. The rms amplitude caused by intrinsic activity in the B band
is about of 0.009 mag. The power of fluctuations at frequencies
below 5 Hz is impossible to evaluate due to large estimation errors.

Fig. 3 indicates the presence of intrinsic activity in the range
$(\sim 0.4 - 3.5)$ Hz in SAO 52355 for an integration time of 0.1 s.
The power excess reaches $\sim 6.8\cdot10^{-5}$ of the total power
in the U band.

Fig. 4 shows the relative power of intrinsic fluctuations in the
range ($\sim 0.5 - 35$) Hz in SAO 52355 for an integration time of
0.01 s. The power excess reaches $2.2 \cdot 10^{-4}$ of the total
power in the V band and corresponds to the rms amplitude of about
0.015 mag. Thus, a new approach based on the theory of count
statistics allowed us to detect activity of SAO 52355 on short
time-scales. The actual value of variability caused by atmospheric
scintillation was determined from measurements of a reference star.
The intrinsic activity of the star was found as a difference between
the observed relative power and that of the atmospheric
scintillations.

We found that the intensity from a variable source on SAO 52355 is
peaked at frequency around about 0.4 Hz, spanning the range up to 35
Hz. The relative power of fluctuations reaches $(6.8 -
22)\cdot10^{-5}$ in the UBV bands. It appears that the observed
variability patterns in SAO 52355 can be related to an ensemble of
microflares with a duration ranging from tenths to a few of seconds.
The energy output of the ensemble-average microflares can be
estimated roughly as $E \approx 4\cdot 10^{-4}$ of the stellar
luminosity. The observations produced by satellite ROSAT had shown
high X-ray luminosity of SAO 52355 that points to presence of a
high-temperature, $T > 10^{7}$ K plasma, suggestive of intense
flaring activity.

On  May 30-31,  2010, we obtained 200 low-resolution grating
spectrograms of SAO 52355. For SAO 52355 we used  GSC 3226 640 as
the reference star, Johnson V magnitude: 10.69. Data inferred from
the Tycho catalog: BT magnitude $11.018 \pm 0.032$, VT magnitude
$10.726 \pm 0.036$. Johnson B-V color, computed from BT and VT:
0.270.

Computed averaged spectra of SAO 52355 and its reference star GSC
3226 640 acquired with the Zeiss-600 telescope are shown in Fig 5.
In Fig 6. the relative power of variations in grating spectra of SAO
52355 and its reference star GSC 3226 640 are shown. Note the
"smooth" spectral energy distribution in the power spectra of the
reference star and the "emission" features at wavelengths of the
Balmer lines and the CaII H, K lines, as well as at a wavelength of
the Balmer jump $\lambda \,\, 3700-3800\,{\AA} $ in SAO 52355. From
the power spectrum data one can find that variations in the
intensities of the CaII H, K and $H_{\gamma}$ lines are 3.2\% and
1.5\%, respectively.

The emission features that identify the hydrogen lines and the CaII
H, K lines, prove chromospheric activity of a star. Thus, the
spectral observations provide an additional argument in favor of the
chromospheric activity of the giant SAO 52355.

\subsection{II Peg}

Further we illustrate the high-frequency activity of II Peg as
resulted from observations at the 2-m telescope at Peak Terskol in
August 2011 with a high-speed photometer. The light curves of II Peg
and its comparison star do not suggest obviously that these stars
are active in the subsecond range (Fig. 7). Fig. 8 indicates,
nevertheless, the presence of intrinsic activity in the range $(\sim
0.4 - 1.5)$ Hz in II Peg for an integration time of 0.1 s. The power
excess reaches $\sim 8.8\cdot10^{-6}$ of the total power in the U
band and corresponds to the rms amplitude of about 0.003 mag.
Variability in the V band is not detected above $\sim 5\cdot10^{-7}$
in almost all frequencies (Fig 9).

\subsection{BD +15 3364}

Fig. 10 indicates as well the presence of intrinsic activity in BD
+153364 in the frequency range $(\sim 0.4 - 1.5)$ Hz. That follows
from high-speed observations with an integration time of 0.1 s at
the 2-m telescope in the B band.

\section{Conclusion}

Photometric monitoring of three chromospherically active stars, BD
+15 3364, II Peg, and SAO 52355, reveal high-frequency variations in
the UBV bands at subsecond range. Intensity variations are found
peaked at frequency around about 0.5 Hz, spanning the range up to
1.5 Hz for BD +15 3364, II Peg, and up to about 35 Hz for SAO 52355.
The relative power of fluctuations reaches $(10^{-3.7} - 10^{-4.2})$
in the UBV bands. Spectroscopic monitoring of SAO 52355 showed
variations of intensity in the Balmer lines and in the CaII H, K
lines at time intervals ranging from seconds to minutes. From the
power spectrum data one can find that variations in the intensities
of the CaII H, K and $H_{\gamma}$ lines are 3.2\% and 1.5\%,
respectively. Satellite surveys at X-ray indicate high temperature
coronae plasma in many chromospherically active stars.
High-frequency changes, which were found in BD +15 3364, II Peg, and
SAO 52355, suggests the existence of intense microflaring activity
in these stars.


\begin{figure*}
\centering
\resizebox{0.90\hsize}{!}{\includegraphics[angle=000]{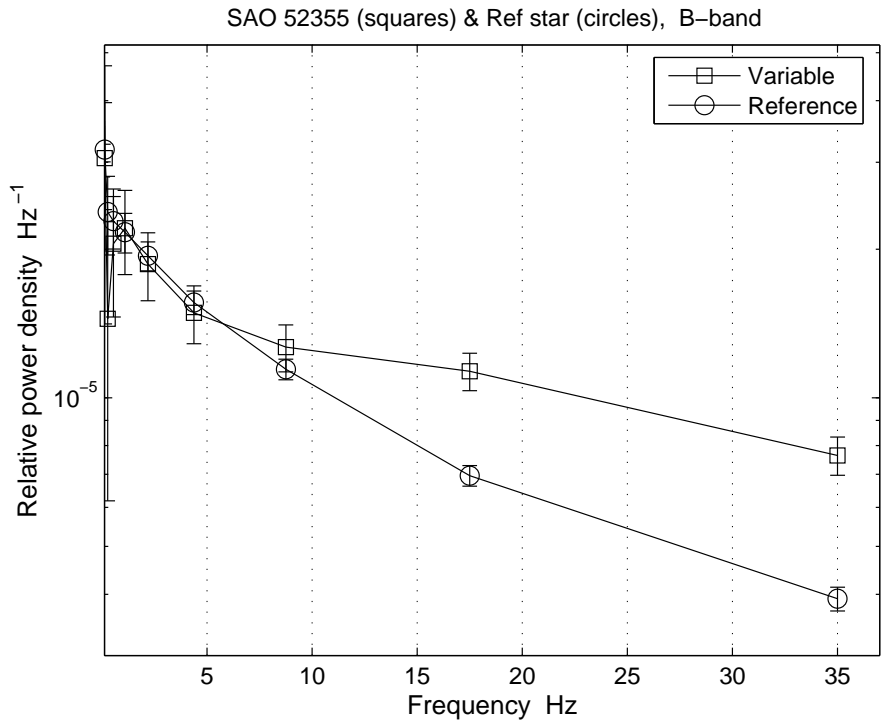}}%
\caption{The relative power density of variations of noise in the B
band for SAO 52355 (denoted by squares) and the comparison star
(denoted by circles) obtained with the 2 m telescope at Peak Terskol
on Aug 28, 2011. The 1-sigma error borders are shown for both
stars.}
\end{figure*}

\begin{figure*}
\centering
\resizebox{0.90\hsize}{!}{\includegraphics[angle=000]{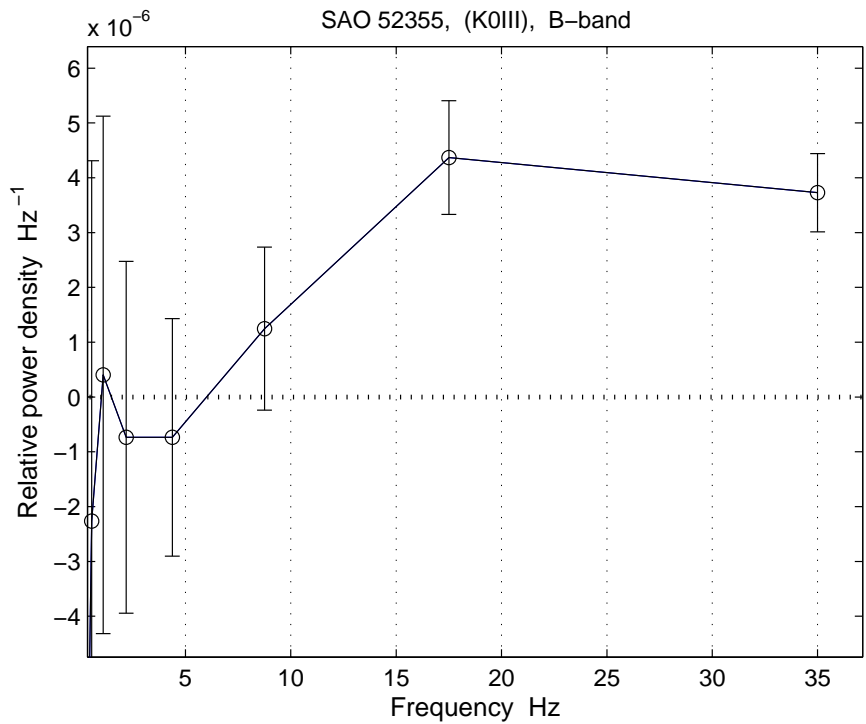}}%
\caption{The relative power density of variations in the B band for
SAO 52355 due to intrinsic activity. The 1-sigma error borders are
shown.}
\end{figure*}

\begin{figure*}
\centering
\resizebox{0.90\hsize}{!}{\includegraphics[angle=000]{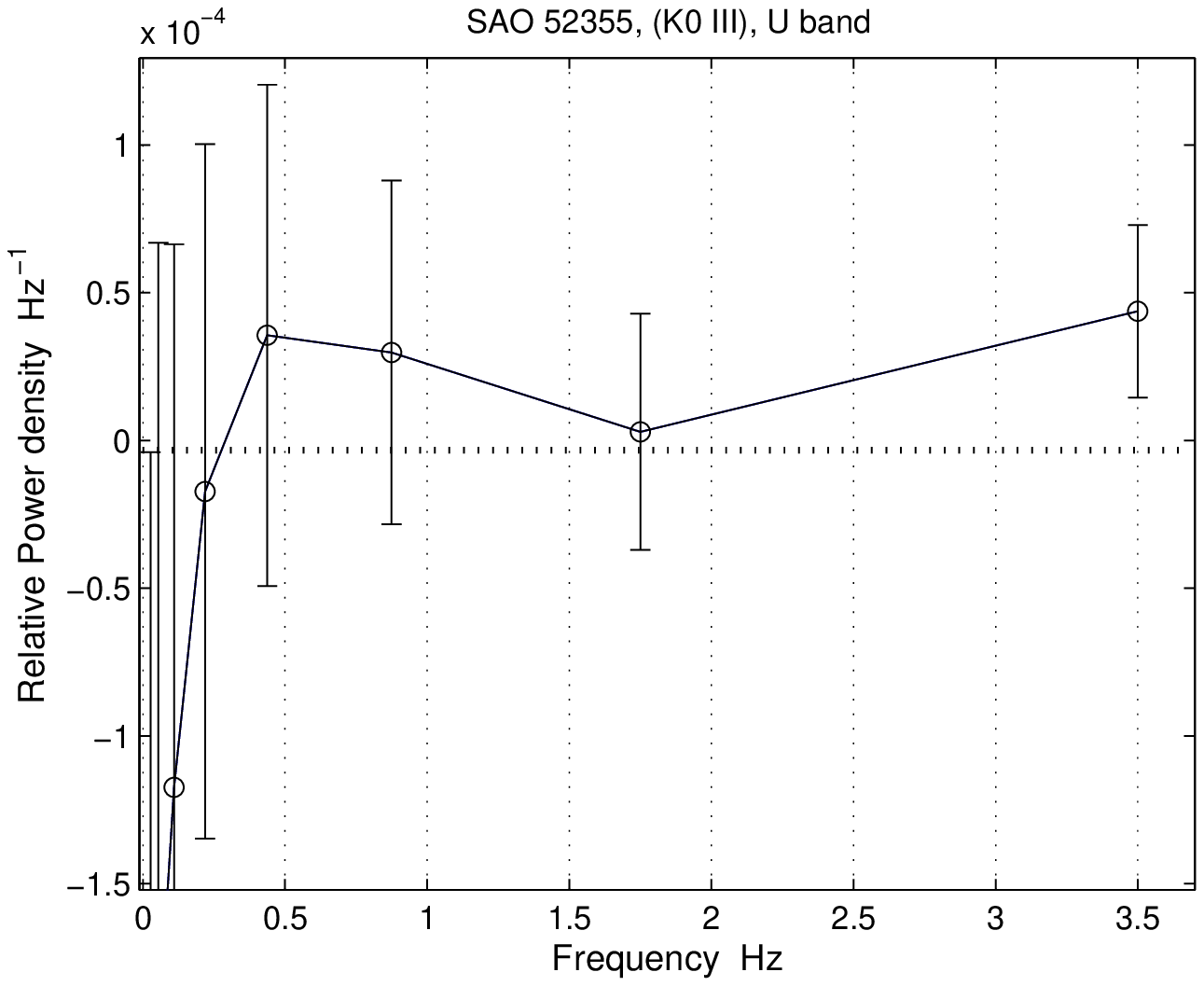}}%
\caption{The relative power density of intrinsic variations in the U
band for SAO 52355. The 1-sigma error borders are shown.}
\end{figure*}

\begin{figure*}
\centering
\resizebox{0.90\hsize}{!}{\includegraphics[angle=000]{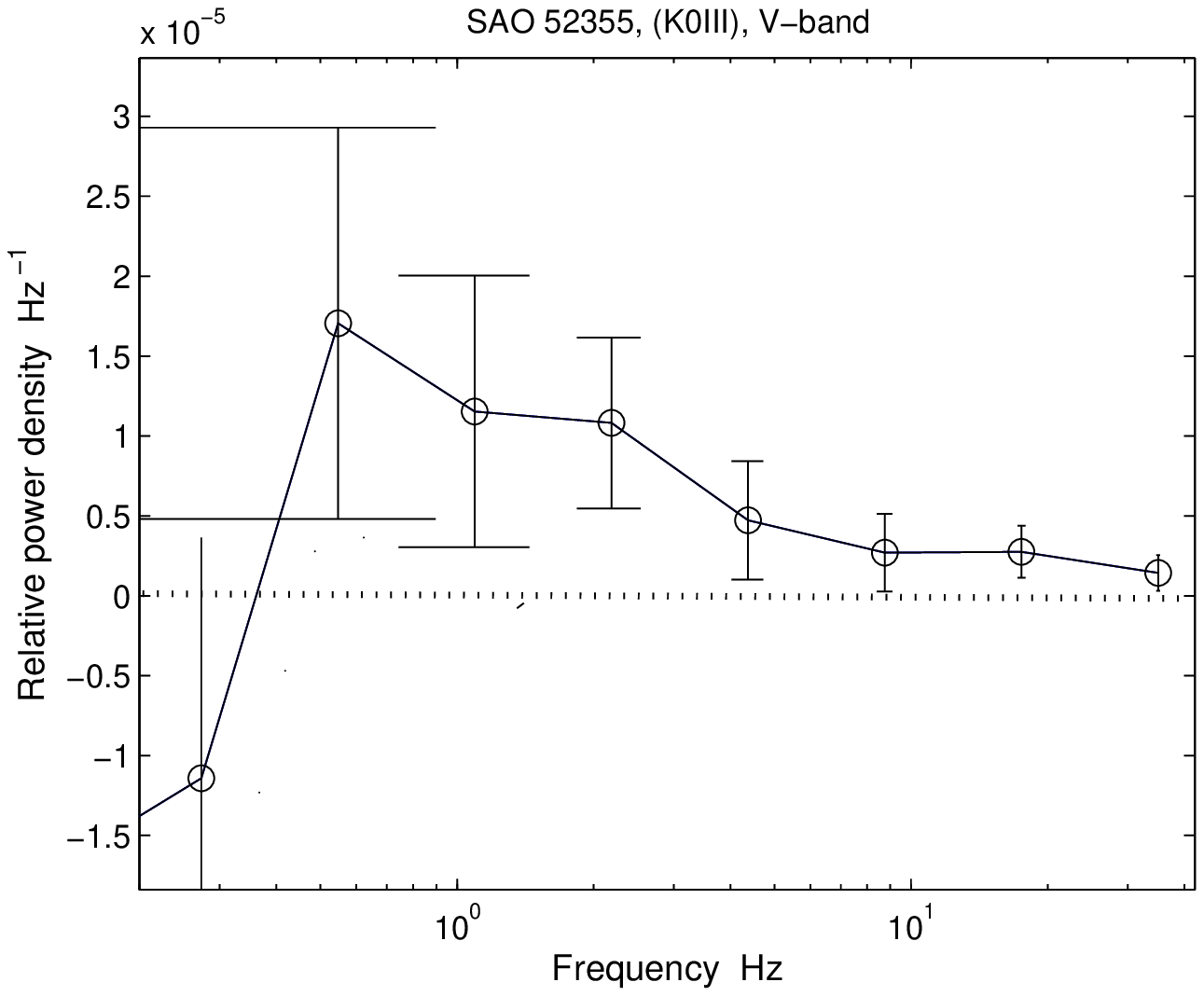}}%
\caption{The relative power density of intrinsic variations in the V
band for SAO 52355. The 1-sigma error borders are shown.}
\end{figure*}

\begin{figure*}
\centering
\resizebox{0.90\hsize}{!}{\includegraphics[angle=000]{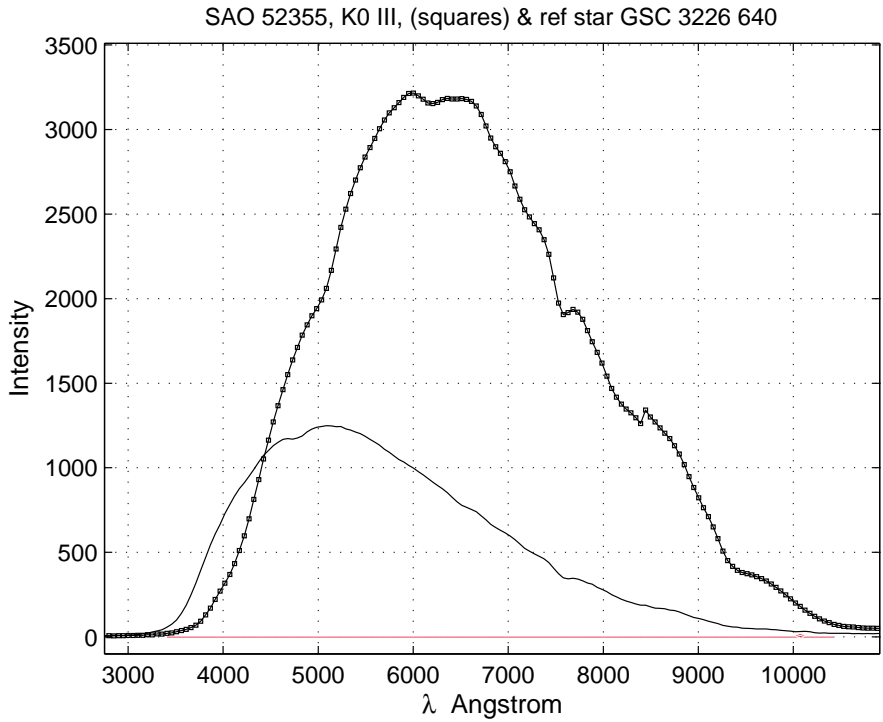}}%
\caption{The averaged raw count rate grating spectra of SAO 52355
(squares) and its reference star  GSC 3226 640 (thin curve).}
\end{figure*}

\begin{figure*}
\centering
\resizebox{0.90\hsize}{!}{\includegraphics[angle=000]{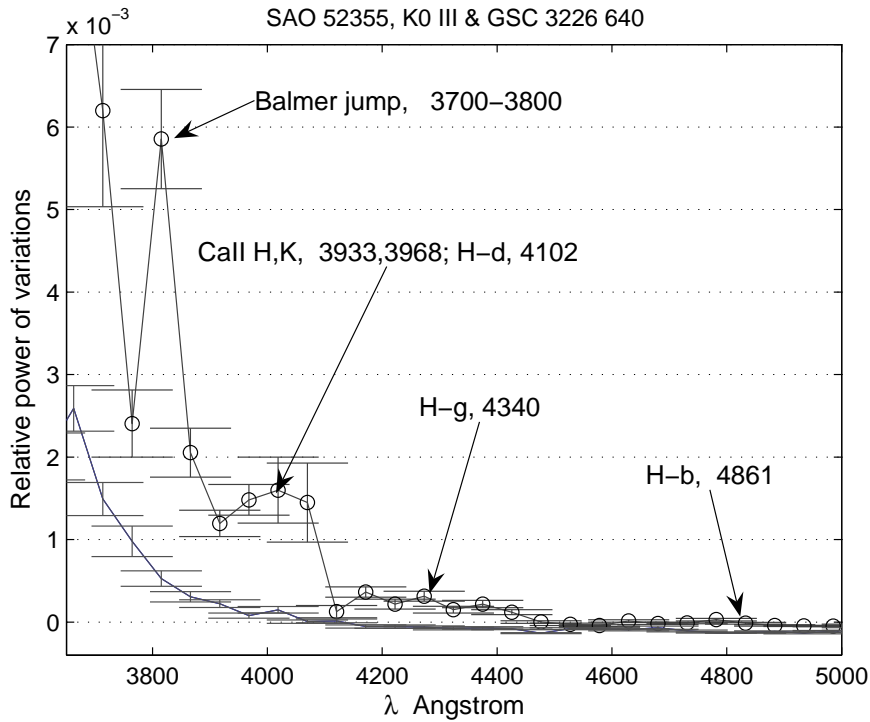}}%
\caption{The relative power of variations in grating spectra of SAO
52355 (circles) and its reference star  GSC 3226 640 (solid curve).1
sigma error bars are shown.}
\end{figure*}

\begin{figure*}
\centering
\resizebox{0.90\hsize}{!}{\includegraphics[angle=000]{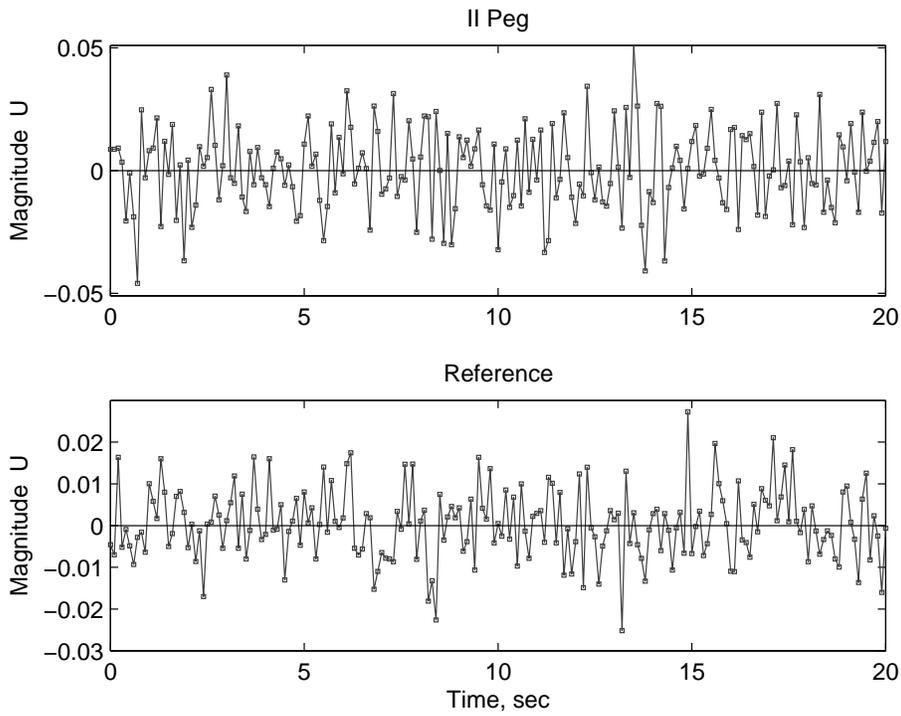}}%
\caption{The light curves of II Peg and a comparison star in the U
band with the sampling time of 0.5 sec.}
\end{figure*}

\begin{figure*}
\centering
\resizebox{0.90\hsize}{!}{\includegraphics[angle=000]{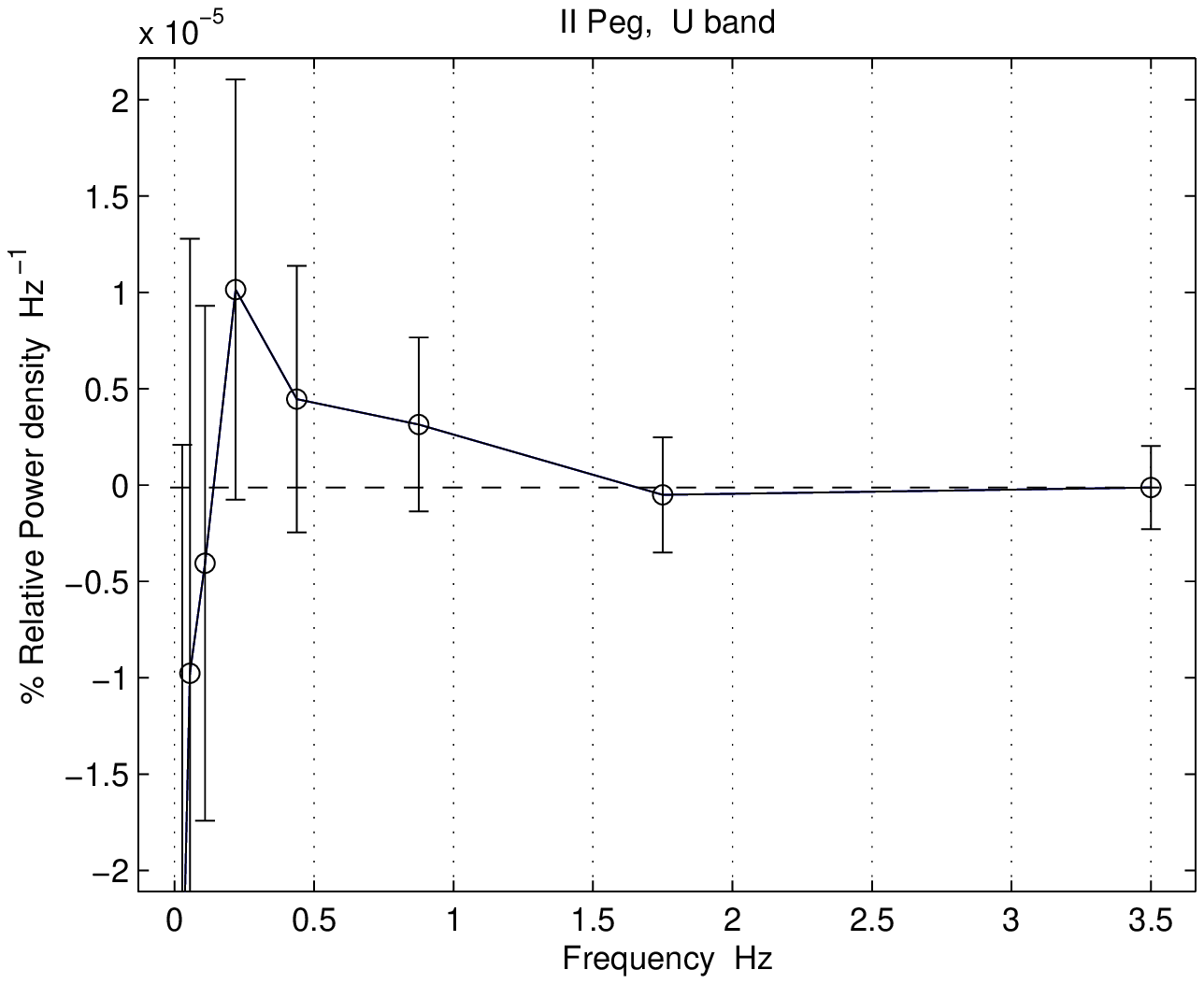}}%
\caption{The relative power density of intrinsic variations in the U
band for II Peg. The 1-sigma error borders are shown}
\end{figure*}

\begin{figure*}
\centering
\resizebox{0.90\hsize}{!}{\includegraphics[angle=000]{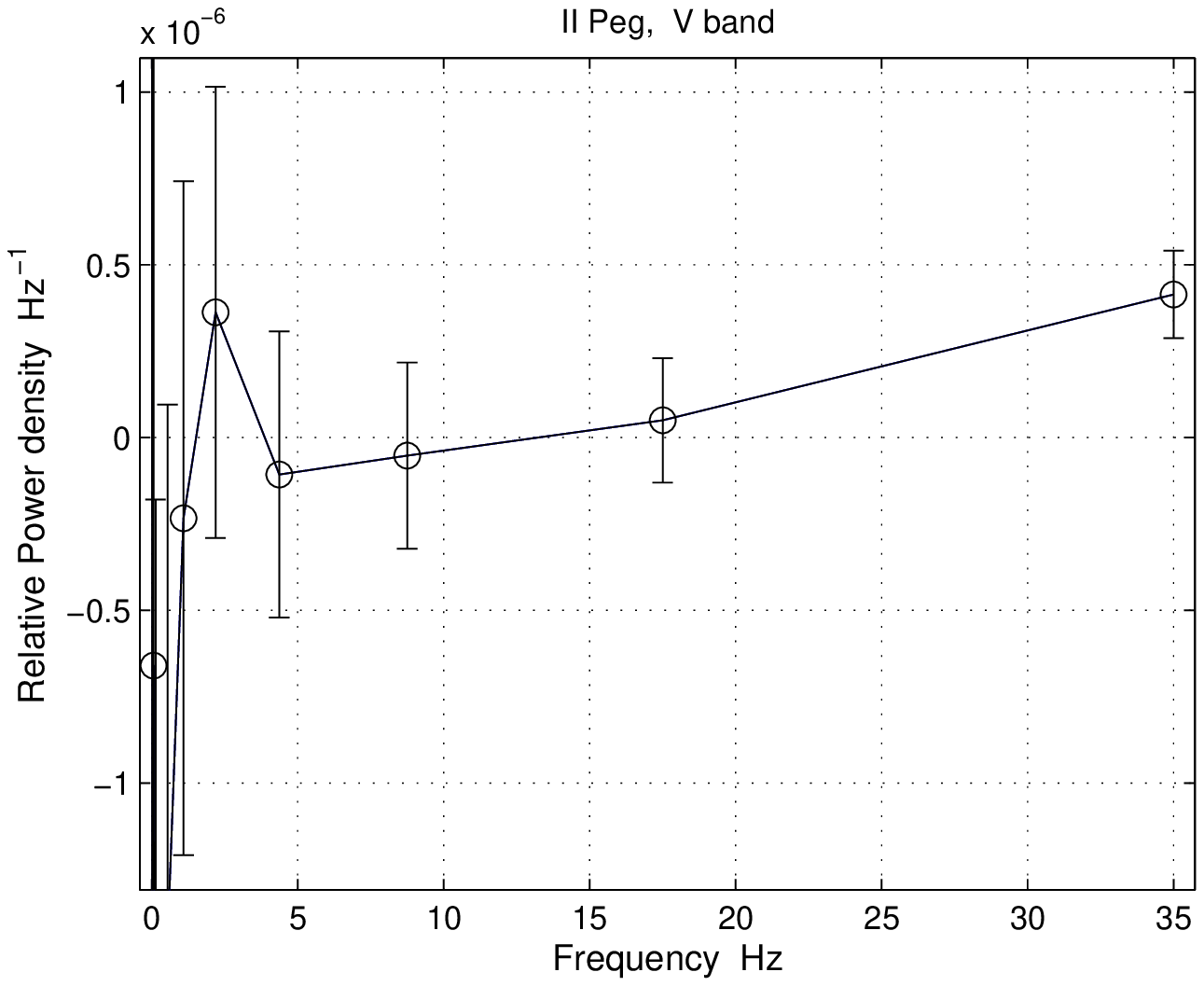}}%
\caption{The relative power density of intrinsic variations in the V
band for II Peg. The 1-sigma error borders are shown}
\end{figure*}

\begin{figure*}
\centering
\resizebox{0.90\hsize}{!}{\includegraphics[angle=000]{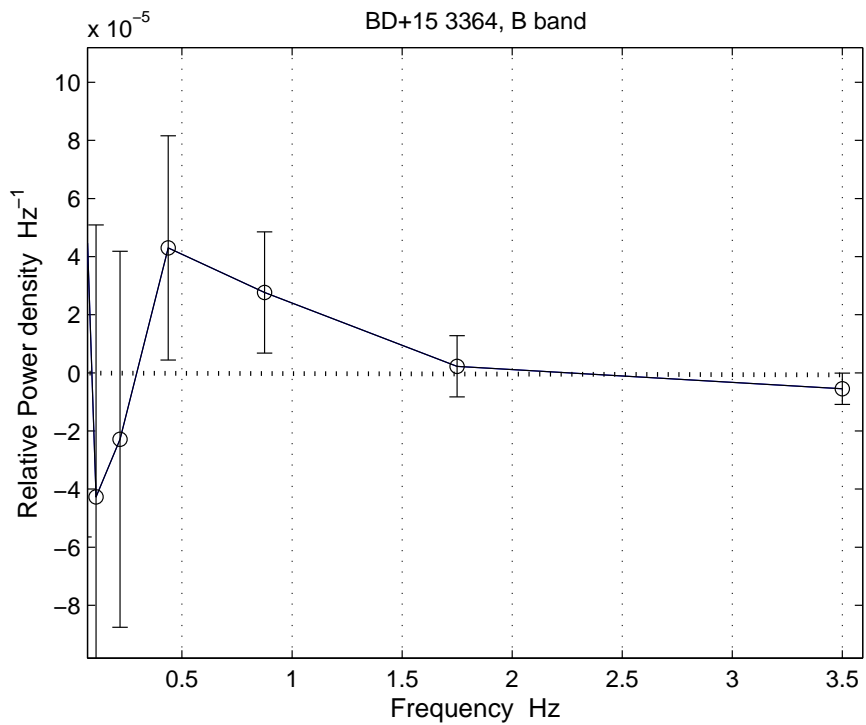}}%
\caption{The relative power density of intrinsic variations in the B
band for BD+15 3364. The 1-sigma error borders are shown}
\end{figure*}


\end{document}